\documentclass{rmaa}
\usepackage{longtable}
\usepackage{pdflscape}

\usepackage{paralist}

\usepackage{psfrag,color}

\usepackage[latin1]{inputenc}

\title{VLT spectroscopic analysis of HH 202. Implications on dust destruction and thermal inhomogeneities.} 

\author{
  Jos\'e N. Esp\'iritu,\altaffilmark{1} 
  Antonio Peimbert,\altaffilmark{1}
  Gloria Delgado-Inglada, \altaffilmark{1}
  and María Teresa Ruiz, \altaffilmark{2}}

\altaffiltext{1}{Instituto de Astronom\'ia, UNAM, M\'exico.}
\altaffiltext{2}{Departamento de Astronom\'ia, Universidad de Chile}

\shortauthor{Esp\'iritu et al.}
\shorttitle{VLT Spectroscopic analysis of HH 202.}

\fulladdresses{
\item J. N. Esp\'iritu, A. Peimbert and G. Delgado-Inglada: Instituto de Astronom\'ia, Universidad Nacional Aut\'onoma de M\'exico, Ap. 70-468, 04510, D.F., M\'exico (jespiritu, antonio, gdelgado@astro.unam.mx)
\item M. T. Ruiz: Departamento de Astronom\'ia, Universidad de Chile, Casilla 36-D, Santiago, Chile (mtruiz@das.uchile.cl)
}
\listofauthors{J. N. Esp\'iritu, A. Peimbert, G. Delgado-Inglada \& M. T. Ruiz.}
\indexauthor{Esp\'iritu, J. N.}
\indexauthor{Peimbert, A.}
\indexauthor{Delgado-Inglada, G.}
\indexauthor{Ruiz, M. T.}

\abstract{We present a long-slit spectroscopic analysis of Herbig-Haro 202 and the surrounding gas of the Orion Nebula using data from the Very Large Telescope.\footnote{Based on observations collected at the European Southern Observatory, Chile, proposal ESO 69.C-0203(A).} Given the characteristics of the Orion Nebula, it is the ideal object to study the mechanisms that play a role in the evolution of \ion{H}{2} regions, notably dust destruction by interstellar shocks, which is a poorly understood subject. The use of long-slit allowed us to determine the spatial variation in its physical conditions and chemical abundances observing a broad area of the Orion Nebula; our results are consistent with those from previous studies albeit with improved uncertainties in some determinations. Special attention is paid to Iron (Fe) and Oxygen (O) abundances, which show a peak at the apex of the shock, allowing us to estimate that 57{\%} of the dust is the destroyed at this position; we also calculate the amount of depletion of oxygen in dust grains, which amounts to 0.126 $\pm$ 0.024 dex. Finally we show that O abundances determined from collisionally excited lines and recombination lines are irreconcilable at the center of the shock unless thermal inhomogeneities are considered along the line of sight in the form of the $t^{2}$ parameter proposed by Peimbert (1967).}

\resumen{}

\addkeyword{ISM: H~II regions}
\addkeyword{Jets and outflows}
\addkeyword{Oxygen depletion}
\addkeyword{Iron depletion}

\begin{document}
\maketitle

\section{Introduction}

The Orion nebula is the brightest \ion{H}{2} region in the night sky. It is considered the standard for studying the chemical composition of \ion{H}{2} regions and the mechanisms that play a role in the evolution of these type of objects. Herbig-Haro (HH) objects  have been studied extensively in molecular clouds, where they can be observed in the infrared, in \ion{H}{2} regions the work has centered mostly on the physical conditions and morphology of these objects \citep[e.g.][]{2001ARA&A..39..403R,2008AJ....136.1566O,2010MNRAS.405.1153S}. Only a few photoionized HH objects have been identified and chemically characterized in \ion{H}{2} regions, notably HH 529 \citep{2006ApJ...644.1006B} and HH 202 \citep{2009MNRAS.395..855M} in the Orion Nebula.

HH 202 is the brightest Herbig-Haro object discovered yet. It was first identified by \citet{1980A&A....85..128C}. Its characteristics allow us to resolve and study the gas flow with high spatial resolution. The parent star has not been identified, however the shock is expanding NW and appears to be related to nearby HH objects HH 529, 203, 204, 528, 269, and 625. The kinematics of  the object are well known; \citet{2008AJ....136.1566O} report a radial velocity between -40 and -60 km/s, while \citet{2009MNRAS.395..855M} conclude that the bulk of emission comes from behind the flow. The object consists of several knots, of which the southern knot (referred to as HH202-S) is the brightest.

HH 202 has been studied previously by \citet{2009MNRAS.395..855M} with the UVES echelle spectrograph of the Very Large Telescope, and \citet{2009MNRAS.394..693M} using integral field spectroscopy. The first work is particularly relevant as it presents an in-depth analysis of the physical conditions and the chemical composition of the shock with high precision. They observed an area of 1.5 $\times$ 2.5 arcsecond$^{2}$ of the sky covering the brightest part of HH 202-S. Their high spectral resolution enabled them to separate the emission from the static gas and the shock. They showed that the heating is due mainly to photoionization by $\theta^{1}$ Ori C, effectively showing that HH 202 can be characterized as an \ion{H}{2} region. They also determined its chemical composition including the presence of thermal inhomogeneities along the line of sight by means of the $t^{2}$ parameter first proposed by \citet{1967ApJ...150..825P}. Finally they calculated the amount of dust destruction and oxygen (O) depletion. 

Although echelle spectrographs provide high spectral resolution, the observed area of the sky is limited to a few arcseconds$^{2}$. For this reason a long-slit study ---which, in the case of FORS 1, allows for the study of a 410\arcsec $\times$ 0.51\arcsec--- of the same area of the sky is excellent to complement and contrast previous results and allow for the spatial exploration of parameters.

The chemical composition of an \ion{H}{2} is usually inferred from its emission spectrum. However, this only represents the gaseous abundance of elements. It is necessary to account for the fraction of a species depleted into dust in order to obtain the total abundance of an element; typically this is reported as a quantity that must be added to the gaseous abundance. Some interstellar shocks are capable of destroying interstellar dust grains if they are energetic enough \citep{2000ApJ...534L..63M}; this phenomenon has been reported in supernova events \citep{2014Natur.511..326G} and Herbig-Haro objects \citep{2009A&A...506..779P, 2009MNRAS.395..855M}.

The case of oxygen trapped in interstellar dust is particularly interesting as it is the third most abundant element in the interstellar medium. The shock velocity of HH 202 is capable of destroying dust grains, making it a good candidate to study the incorporation of oxygen and other elements from the dust phase into the gas phase. \citet{1998MNRAS.295..401E} and \citet{2009MNRAS.395..855M} have estimated the depletion correction for oxygen to be 0.08 dex and 0.12 $\pm$ 0.03 dex respectively. 

In this work, we have conducted an analysis of HH 202 using the long-slit Focal Reducer Low Dispersion Spectrograph 1 (FORS 1) of the Very Large Telescope. In Section \ref{sec.observations} we present our observations and the data processes we used. We perform a spatial analysis of the iron (Fe) and oxygen emissions in Section \ref{sec.FeOemission}, including the spatial variation in abundance across the Orion nebula. In Section \ref{sec.combinedanalysis} we present our results from combining multiple spectra identifying the zones where the shock due to HH 202 is most prominent; electron density and temperature are calculated and ionic and total abundances are presented assuming constant temperature and thermal inhomogeneities. Finally we calculate the oxygen depletion inferred from dust destruction and the conclusion to our analysis in sections \ref{sec.discussion} and \ref{sec.conclusions}.

\section{Observations and data reduction.}\label{sec.observations}

The observations were carried out during the night of September 11, 2002 with FORS 1 at the Very Large Telescope (VLT), in Cerro Paranal, Chile. Data were obtained from three different grism configurations: GRIS-600B+12, GRIS-600R+14 with filter GG435+31, and GRIS-300V+10 with filter GG375+30 (see Table \ref{table.observations}).

\begin{table}\centering
\caption{Journal of observations}
\label{table.observations}
\begin{tabular}{c c c c c}
\toprule
Grism & Filter & $\lambda$ (\AA) & Resolution ($\lambda/\Delta \lambda$) & Exposure time (s) \\ 
\midrule
GRIS-600B+12	& \nodata 	&	3450--5900	&	1300	& 3 $\times$ 60 \\
GRIS-600R+14	& GG435		& 	5250--7450	&	1700	& 5 $\times$ 30 \\
GRIS-300V	& GG375		&	3850--8800	&	700	& 3 $\times$ 20 \\
\bottomrule	
\end{tabular}
\end{table}

\begin{figure}
\includegraphics{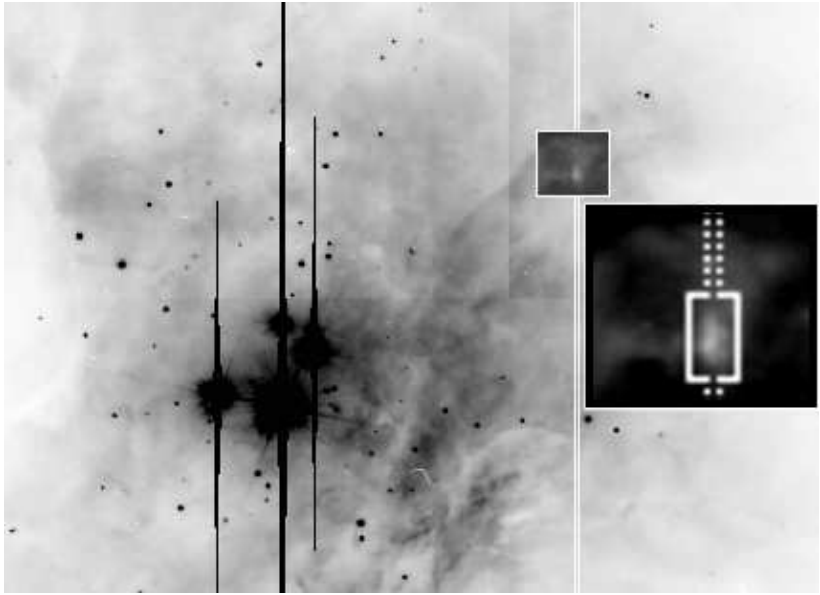}
\caption{Central part of the Orion Nebula. The white vertical lines show the position and width of the slit used. The region inside the white box represents HH202-S. A close-up image of the shock is also shown; the white rectangle encloses the zone with the peak of emission of H$\beta$, [\ion{Fe}{2}] $\lambda$7155, and [\ion{Fe}{3}] $\lambda$4658. North points to the top of the image and East to the left.}
\label{fig.orion}
\end{figure}

An image of the Orion Nebula from our observations can be seen in figure \ref{fig.orion}. The slit was oriented North--South, and the atmospheric dispersion corrector was used to keep the same observed region within the slit regardless of the airmass value. The slit length was 410\arcsec and the width was set to 0.51 \arcsec. This setting was chosen to have the resolution to deblend the [\ion{O}{2}] $\lambda$3726 and $\lambda$3729 emission lines, as well as measuring \ion{O}{2} $\lambda$4642 and $\lambda$4650 with a significant signal to noise ratio with GRIS-600B+12.

The final spectrum was reduced using IRAF\footnote{IRAF is distributed by National Optical Astronomy Observatories, which is operated by the Association of Universities for Research in Astronomy, under cooperative agreement with the National Science Foundation).} following the standard procedure of bias substraction, aperture extraction, flat fielding, wavelength calibration and flux calibration. The standard stars used for this purpose were LTT 2415, LTT 7389, LTT 7987, and EG 21 \citep{1992PASP..104..533H, 1994PASP..106..566H}. The error in flux calibration was estimated to be 1\%.

To analyze the spatial variations of the physical properties and of the chemical abundances 54 extraction windows were defined. Windows North and South of HH202-S span 50 pixels (10 \farcs{}) each, whereas those covering the object are 3 pixels (0.6 \farcs{}) long  each. This treatment of the data allowed us to establish the composition of the Herbig-Haro object and compare it directly with the surrounding gas of the Orion nebula. The apex of HH202-S ---that is, the region where the shock is strongest--- has coordinates (J2000.0) $\alpha$ = 05$^{h}$35$^{m}$11$^{s}$.6, $\delta$ = -5\arcdeg 22\arcmin 56\farcs{}. The former was determined from the peak of emission of [\ion{Fe}{2}] $\lambda$7155; [\ion{Fe}{3}] $\lambda$4658;  \ion{O}{2} $\lambda \lambda$4640 and 4652; and H$\alpha$.

\section{Spatial analysis}\label{sec.FeOemission}

\begin{figure}
\centering
\includegraphics[width=\textwidth]{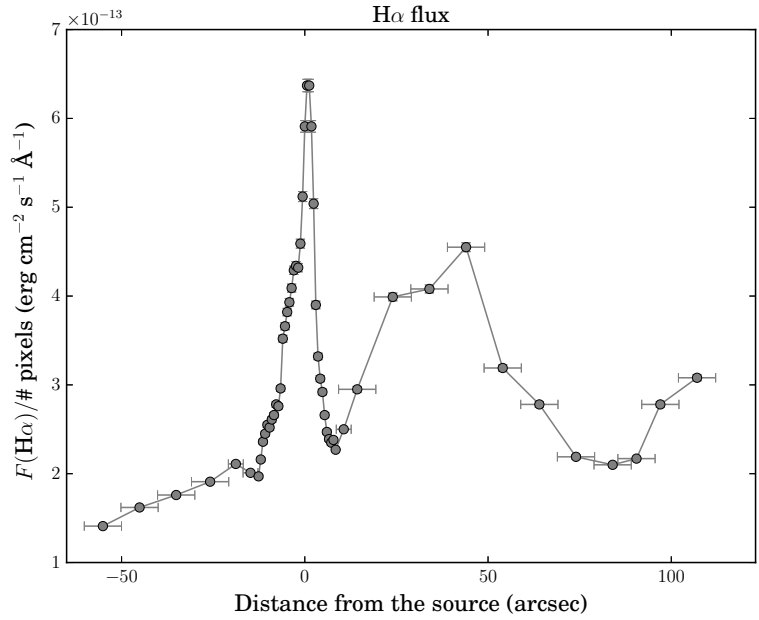}
\caption{H$\alpha$ flux across the Orion Nebula. The zero mark was determined from the peak of [\ion{Fe}{3}] emission with approximate coordinates $\alpha$ = 05$^{h}$35$^{m}$11$^{s}$.6 and $\delta$ = -5\arcdeg,22\arcmin,56\farcs{}2 (2000). North is to the left of the zero mark and South is to the right.}
\label{fig.Halpha_flux}
\end{figure}

\begin{figure}
\centering
\includegraphics[width=\textwidth]{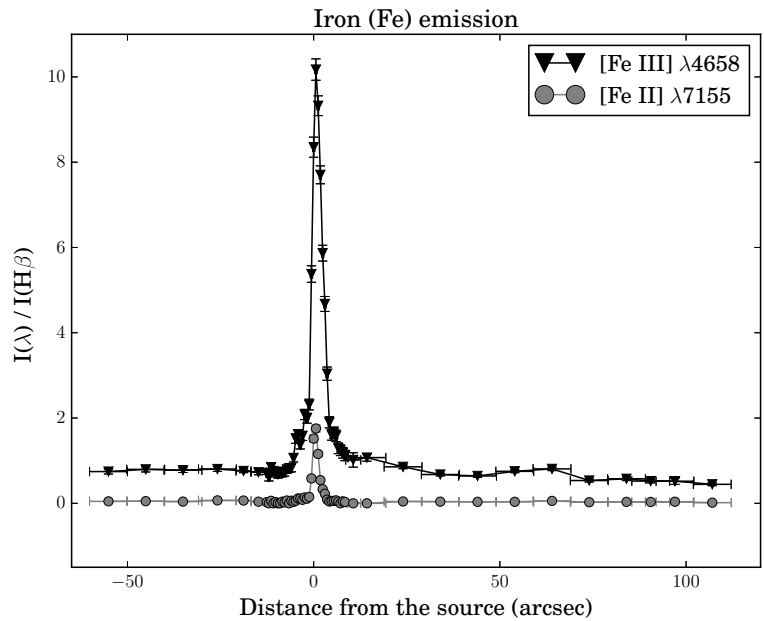}
\caption{Dereddened emission line intensities for [\ion{Fe}{2}] $\lambda$7155 and [\ion{Fe}{3}] $\lambda$4658 across the Orion Nebula.}
\label{fig.iron_emission}
\end{figure}

We performed an analysis of the flux of a set of emission lines on all 54 windows: the Balmer series up to H9; [\ion{Fe}{2}] $\lambda$ 7155; [\ion{Fe}{3}] $\lambda$ 4658; and \ion{O}{2} $\lambda$$\lambda$4640 and 4652.  The flux of the emission lines was determined by integrating between two points over the local continuum estimated by eye. This was done using the SPLOT routine of IRAF. A Gaussian profile was fitted to the lines that were blended together. The results for H$\alpha$ are presented in Figure \ref{fig.Halpha_flux} showing a peak that coincides perfectly with the brightest section of the slit covering the object (Figure \ref{fig.orion}). This peak also coincides with the peak of the iron emission lines presented in Figure \ref{fig.iron_emission}; indicating the center of the shock.

We tested the extinction laws of \citet{1979MNRAS.187P..73S}, \citet{1989ApJ...345..245C}, and \citet{1970BOTT....5..229C}. The logarithmic extinction correction for H$\beta$, $C$(H$\beta$), and the underlying absorption were fitted simultaneously to the theoretical ratios. The theoretical intensity ratios for the Balmer emission lines were calculated using INTRAT by \citet{1995MNRAS.272...41S} considering a constant electron temperature $T_{e} = 9 000$ K, and an electronic density $n_{e} = 5 000$ cm$^{-3}$; there was no need to modify these values since hydrogen lines are nearly independent from temperature and density. The underlying absorption ratios for the Balmer and helium emission lines were taken from Table 2 of \citet{2012ApJ...746..115P}. The most suitable values for $C$(H$\beta$) and the underlying absorption in H$\beta$, EW$_{\mathrm{abs}}$(H$\beta$), were found by reducing the quadratic discrepancies between the theoretical and measured H lines in units of the expected error, $\chi^{2}$. The extinction law by \citet{1970BOTT....5..229C} delivered the most satisfying results and is the one we adopted for the rest of this work. The fluxes were normalized with respect to the whole Balmer decrement, meaning that the value of $I$(H$\beta$) was allowed to deviate slightly from 100. 

The emission line intensities for [\ion{Fe}{2}] $\lambda$7155 and [\ion{Fe}{3}] $\lambda$4658 are presented in Figure \ref{fig.iron_emission}. We can see how the intensities of both lines increase by an order of magnitude for a region about 4 arcseconds in length, this increase in intensity cannot be explained by differences in temperature or density, it must be caused then by a great increase in the amount of iron in the gaseous phase: indicating dust destruction on a considerable scale. We will define the zero point in our coordinates as the one corresponding to maximum [\ion{Fe}{2}] and [\ion{Fe}{3}] intensities. 

\begin{figure}
\centering
\includegraphics[bb=30 411 662 680,clip,width=\textwidth]{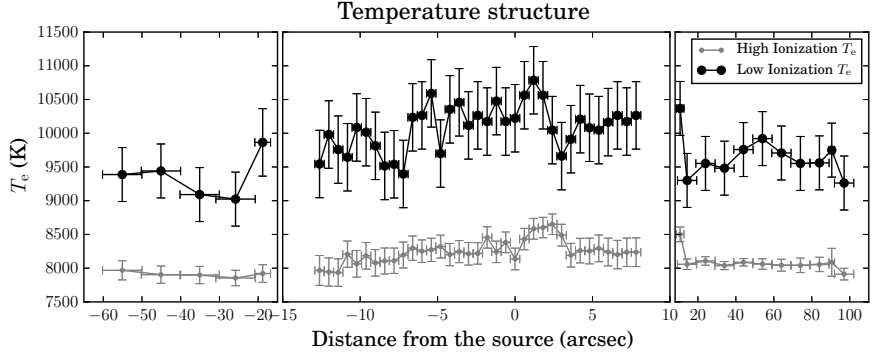}
\caption{The low-ionization electron temperature corresponds to [\ion{N}{2}] $\lambda$5755/$\lambda$6584; the high-ionization temperature was calculated using [\ion{O}{3}] $\lambda$4363/$\lambda$5007.}
\label{fig.temperature}
\end{figure}

\begin{figure}
\centering
\includegraphics[bb=29 412 668 679,clip,width=\textwidth]{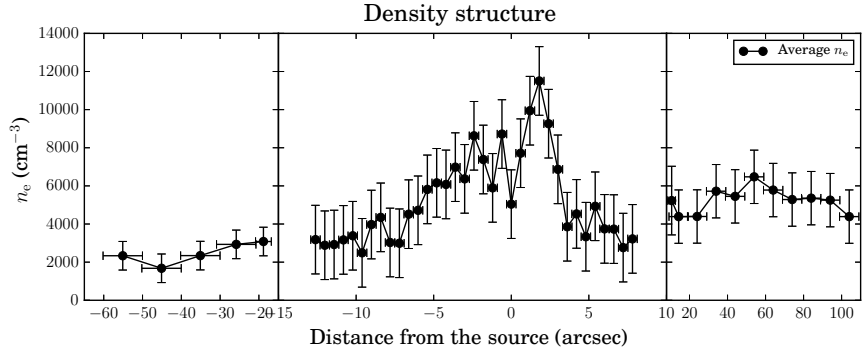}
\caption{Electron density shown is an average of the [\ion{S}{2}] $\lambda$$\lambda$ 6716/6731 and [\ion{Cl}{3}] $\lambda$$\lambda$ 5518/5538 diagnostics}.
\label{fig.density}
\end{figure}

Electron temperatures, $T_{\mathrm{e}}$, and densities, $n_{\mathrm{e}}$, across the area of the Orion Nebula covered by the slit are presented in Figures \ref{fig.temperature} and \ref{fig.density}. The [\ion{N}{2}] $\lambda$6584/  $\lambda$5755 and [\ion{O}{3}]  $\lambda$4363/$\lambda$5007 ratios were used to determine the low- and high-ionization temperatures. For electron density, we computed the average of the [\ion{S}{2}] $\lambda$6716/$\lambda$6731, and [\ion{Cl}{3}] $\lambda$5517/$\lambda$5537 ratios since the uncertainties associated with the latter are of a considerable size. The references for the atomic data set used to compute physical conditions and chemical abundances are presented in Table \ref{table.atomicdata}.

The physical conditions reported here were obtained using PyNeb \citep{2015A&A...573A..42L}, by identifying the intersection of the corresponding temperature and density diagnostics. Our results for the North and South zones away from the shock agree with previous determinations made by \citet{2003MNRAS.340..362R}, \citet{2004MNRAS.355..229E} and \citet{2009MNRAS.395..855M}. In the case of the shocked spectra, our results overlap with the upper limit reported for the [\ion{N}{2}] electron temperature by \citet{2004MNRAS.355..229E}, also with the lower limits for the [\ion{Cl}{3}] and [\ion{O}{2}] densities; our reported [\ion{O}{3}] temperature is about 300 K higher however. We attribute this difference to the fact that we are not observing the same volume of gas; also, differences in calibration and the extinction law used may cause these minor disparities; however, this does not have major implications on the chemical abundances since the dependency of an emission line intensity with $n_{\mathrm{e}}$ is minimal; moreover, for recombination lines the dependency with temperature is negligible.

\begin{figure}
\centering
\includegraphics[bb=52 502 802 820,clip,width=\textwidth]{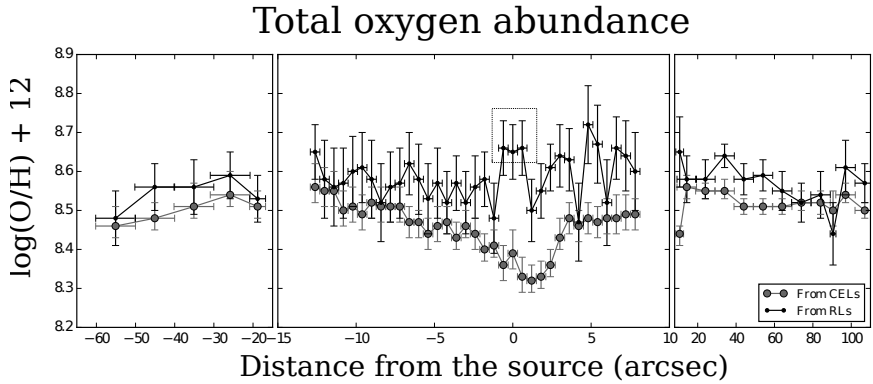}
\caption{Total O/H ratio computed using CELs and \ion{O}{2} RLs (see text). Note that the difference is maximum at the apex of the shock.}
\label{fig.oxigeno_total}
\end{figure}

We computed the total abundance for oxygen in two ways: from Collisionally Excited Lines (CELs) and Recombination Lines (RLs). The abundance from CELs, O$_{\mathrm{CEL}}$, is the sum of the O$^{+}$ and O$^{2+}$  ionic abundances, obtained from [\ion{O}{3}] $\lambda$5007 and [\ion{O}{2}] $\lambda$3726+29 respectively. For the RL oxygen abundance we have used multiplet 1 of \ion{O}{2} to determine O$^{2+}/\mathrm{H}^{+}$. The intensity of multiplet 1 of \ion{O}{2} is the sum of eight lines, of which we only detected four blended in pairs as $\lambda$$\lambda$4639 + 42 and $\lambda$$\lambda$4649 + 51. The total intensity was estimated considering the dependence on density and temperature of the lines, according to the work of \citet{2010ApJ...724..791P}. The effective recombination coeficients were taken from \citet{1994A&A...282..999S} for case B, assuming $n_{\mathrm{e}}= 10,000$ cm$^{-3}$. Although \ion{O}{I} lines are present in our spectra, they are contaminated by telluric emission, making them unreliable for calculating O$^{+}$;  to account for O$^{+}$ we have assumed the following relation between O$^{+}$ and O$^{2+}$:
\begin{equation}
\left[ \frac{\mathrm{O}}{\mathrm{H}} \right]_{\mathrm{RL}} = \left[ \frac{\mathrm{O}^{2+} + \mathrm{O}^{+}}{\mathrm{O}^{+}} \right]_{\mathrm{CEL}}   \times \left[ \frac{\mathrm{O}^{2+}}{\mathrm{H}^{+}}  \right]_{\mathrm{RL}}.
\end{equation}
\citet{2004MNRAS.355..229E} also favor the former procedure.  Oxygen abundances derived from RLs and CELs are presented in Figure \ref{fig.oxigeno_total}. While small, there is a significant difference between the oxygen abundance near the apex of HH202-S compared with the surrounding ---presumed static--- gas; it is also evident from Figure \ref{fig.oxigeno_total} that  O$_{\mathrm{CEL}}$ and O$_{\mathrm{RL}}$ are irreconcilable in the line of sight of the shock.

It is known that oxygen is present in interstellar dust grains in the form of water ice and metallic compounds such as FeO, CaO and MgO. Theoretical and empirical studies have shown that dust can be destroyed by grain-grain collisions in interstellar shocks ---a process known as sputtering--- thus reincorporating refractory elements into the diffuse gas,  however these studies have been carried out mostly in molecular clouds (see, for example \citet{2009A&A...506..779P} and references therein) and supernova remnants \citep{2014Natur.511..326G}, \citep{2000ApJ...534L..63M}. Dust composition in \ion{H}{2} regions is known to be different from that found in molecular clouds due to photo-evaporation of ice molecules by UV radiation; in any case the relation between dust destruction and shock velocity is not entirely clear. The only work to study dust destruction in a Herbig-Haro object in an \ion{H}{2} region and its effect on oxygen was made by \citet{2009MNRAS.395..855M} who also examined HH202-S, finding an increase in oxygen, iron and magnesium abundance at the shock; showing the presence of dust destruction in the aftermath of moderate shockwaves. This effect is also present in our observations.

There has been a long debate on the magnitude of thermal inhomogeneities in \ion{H}{2} regions and on their effect on the determination of chemical abundances \citep{1969BOTT....5....3P, 2011A&A...526A..48S, 2012ApJ...746..115P}. Regardless of the typical effect on \ion{H}{2} regions, an interstellar shock is clearly a case where non-negligible temperature variations are expected. Given that RLs are not affected by temperature variations to the same degree as CELs, we favored abundance determinations done with RLs for the analysis of dust destruction.

\begin{table}
\caption{Atomic data set}
\label{table.atomicdata}
\begin{tabular}{c c c}
\toprule
Ion & Transition probabilities & Collision strengths \\
\midrule
N$^{+}$ & \citet{1996wiese} \citet{1997galavis} & \citet{2011tayal} \\
O$^{+}$ & \citet{1996wiese}, \citet{2006MNRAS.366L...6P} & \citet{2007ApJS..171..331T} \\
O$^{2+}$ & \citet{1996wiese}, \citet{2000MNRAS.312..813S} & \citet{1999ApJS..123..311A} \\
S$^{+}$ & \citet{1982MNRAS.198..127M} & \citet{2010ApJS..188...32T} \\
S$^{2+}$ & \citet{1982MNRAS.199.1025M} & \citet{1999ApJ...526..544T} \\
Cl$^{2+}$ & \citet{1983IAUS..103..143M} & \citet{1989butler} \\
Ar$^{2+}$ & \citet{1983IAUS..103..143M} & \citet{1995galavis} \\
Fe$^{2+}$ & \citet{1996quinet}, \citet{2000johan} & \citet{1996zhang} \\
Ni$^{2+}$ & \citet{2001bautista} & \citet{2001bautista} \\
\bottomrule
\end{tabular}
\end{table}
\section{Analysis from combined spectra}\label{sec.combinedanalysis}

In order to enhance the signal to noise ratio and reduce the bias produced by measuring very weak lines we decided to combine the three spectra with the highest Fe and O abundance (the ones with maximum dust destruction), which we will call the strongly shocked zone (SS), represented in Figure \ref{fig.oxigeno_total} with a box at the zero mark. To represent the static gas we chose two regions: one averaging 4 windows 20 arcseconds south of HH 202-S (South Zone), and one averaging 4 regions 20 arcseconds north of HH 202-S (North Zone). Finally we also combined the spectra of four weakly shocked zones (WS). The resulting spectra were thoroughly studied to derive most of the conclusion of this work. 

The dereddened fluxes for H$\beta$ corresponding to the extinction law of \citet{1970BOTT....5..229C} are presented at the bottom of Table \ref{longtable.emission_lines}.

The emission line intensities for the four combined spectra covering Northern and Southern zones of the Orion Nebula as well as HH202-S are presented in Table \ref{longtable.emission_lines} in columns 4--11. Column 1 shows the laboratory wavelength $\lambda$ for air; column 2 presents the identification for each line based on the work by \citet{2004MNRAS.355..229E} and the Atomic Line List v2.04\footnote{The Atomic Line List is maintained by Peter van Hoof: http://www.pa.uky.edu/~peter/atomic/.}; column 3 presents the value of $f(\lambda)$ for each line. Overall, we have identified 169 different emission lines in our combined spectra; the Strongly Shocked Zone had the most emission lines, with 159, including several additional Cr and Fe lines.

From Table \ref{longtable.emission_lines} we can ascertain that dust is being destroyed by the shock front. Comparing the intensity of the iron emission lines in the strongly shocked zone with an average for the North and South zones we find that all of them are brighter at the apex of HH202-S; particularly, [\ion{Fe}{2}] $\lambda$7155 is 26 times more  intense  at the shock and [\ion{Fe}{3}] $\lambda$4658 is 13 times brighter. The increase in the gaseous abundance at the shock is due to the incorporation of this iron by the destruction of dust grains.

\begin{figure}[h]
\includegraphics[width=\textwidth]{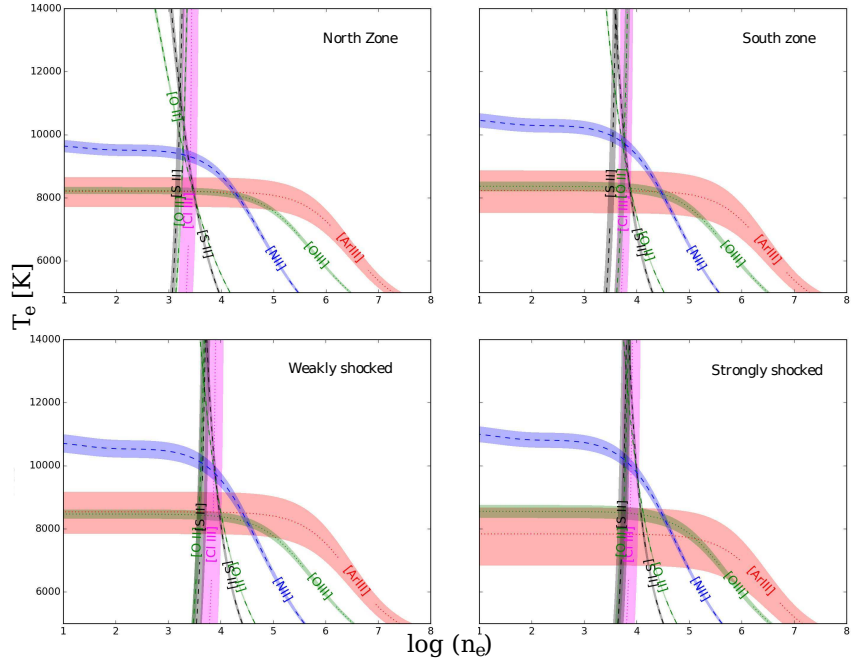}
\caption{Electron temperature and density diagnostics for the four analyzed zones of the Orion Nebula.}
\label{fig.diagnostics}
\end{figure}

We computed diagnostics for $T_{\mathrm{e}}$ and $n_{\mathrm{e}}$ using PyNeb. For high ionization we used: [\ion{O}{3}] $\lambda$4363/$\lambda$5007, and [\ion{Ar}{3}] $\lambda$5192/$\lambda$7136 for $T_{e}$; and [\ion{Cl}{3}] $\lambda$5518/$\lambda$5538 for $n_{e}$. For low ionization we computed: [\ion{N}{2}]  $\lambda$5755/$\lambda$$\lambda$6548 + 84, [\ion{O}{2}] $\lambda$$\lambda$3726 + 29 /$\lambda$$\lambda$7319 + 30, [\ion{S}{2}] $\lambda$$\lambda$4069 + 76/$\lambda$$\lambda$6716+31 for $T_{e}$; and [\ion{O}{2}] $\lambda$3726/$\lambda$3729, [\ion{S}{2}] $\lambda$6716/$\lambda$6731 for $n_{e}$. These diagnostics are presented in Figure \ref{fig.diagnostics}. The high ionization temperature and density were determined from the intersection of the [\ion{O}{3}] and [\ion{Cl}{3}] diagnostics. For low ionization, the physical conditions were determined graphically from the available diagnostics (see the aforementioned figures) by establishing the midpoint between the [\ion{N}{2}], [\ion{O}{2}], and [\ion{S}{2}] lines. Table \ref{table.temden} presents the specific physical conditions for each diagnostic as well as the adopted $T_{\mathrm{e}}$ and $n_{\mathrm{e}}$.

We have followed the formalism developed by \citet{1967ApJ...150..825P} to account for thermal inhomogeneities in the temperature structure of the nebula along the line of sight. This approach establishes an average temperature, $T_{0}$ and the mean square temperature inhomogeneities, $t^{2}$, defined as:
\begin{equation}
T_{0}(\mathrm{ion}) = \frac{\int T_{e}(\mathbf{r})n_{e}(\mathbf{r}) n_{\mathrm{ion}} (\mathbf{r}) dV }{\int n_{e} (\mathbf{r})  n_{\mathrm{ion}} (\mathbf{r}) dV},
\end{equation}
\begin{equation}
t^{2}(\mathrm{ion}) = \frac{ \int (T_{e} - T_{0})^{2} n_{e}(\mathbf{r}) n_{\mathrm{ion}}(\mathbf{r}) dV }{ T_{0}^{2} \int n_{e}(\mathbf{r}) n_{\mathrm{ion}} (\mathbf{r}) dV}
\end{equation}

For the case of O$^{2+}$ we can derive the following equation \citep{2004ApJS..150..431P}:
\begin{equation}
T_{4363/5007} = T_{0} \left[ (1 + \frac{t^{2}}{2} \left( \frac{91300 \mathrm{K}}{T_{0}} - 3 \right) \right]. 
\label{eq.temperature}
\end{equation}
A similar equation can be derived for the $T_{0}$ of the low-ionization species.

Once $T_{0}$ and $t^{2}$ have been determined, Equation \ref{eq.temperature} is implemented to calculate the O$^{2+}$ abundance \citep{1969BOTT....5....3P, 2004MNRAS.355..229E}.

\scriptsize

\begin{changemargin}{-2cm}{-1cm}
\centering
\setlength{\tabnotewidth}{\columnwidth}
  \tablecols{11}
\begin{longtable}{l c c r r r r r r r r}
\toprule
\endfirsthead
\endhead
\caption[]{\centering\normalsize{\uppercase{List of emission line intensities for HH202-S and the Orion Nebula\tabnotemark{a}.}}}
\label{longtable.emission_lines}\\
\toprule
\multicolumn{3}{c}{ } & \multicolumn{2}{c}{North zone} & \multicolumn{2}{c}{WS zone} & \multicolumn{2}{c}{SS zone} & \multicolumn{2}{c}{South zone}\\
\cmidrule(l){4-5} 
\cmidrule(l){6-7}
\cmidrule(l){8-9}
\cmidrule(l){10-11} 
$\lambda$ & Ion 		& $f(\lambda)$ 	&\multicolumn{1}{c}{$I$} 	&\multicolumn{1}{c}{{\%}err}& \multicolumn{1}{c}{$I$}&\multicolumn{1}{c}{{\%}err}& \multicolumn{1}{c}{$I$}&\multicolumn{1}{c}{{\%}err}& \multicolumn{1}{c}{$I$} 	& \multicolumn{1}{c}{{\%}err }\\ \hline
\endfirsthead
\caption[]{continued.}\\
\midrule
\midrule
\multicolumn{3}{c}{ } & \multicolumn{2}{c}{North zone} & \multicolumn{2}{c}{WS zone} & \multicolumn{2}{c}{SS zone} & \multicolumn{2}{c}{South zone}\\
\cmidrule(l){4-5} 
\cmidrule(l){6-7}
\cmidrule(l){8-9}
\cmidrule(l){10-11} 
$\lambda$ & Ion 		& $f(\lambda)$ 	& \multicolumn{1}{c}{$I(\lambda)$} 	&\multicolumn{1}{c}{{\%}err}& \multicolumn{1}{c}{$I(\lambda)$}&\multicolumn{1}{c}{{\%}err}& \multicolumn{1}{c}{$I(\lambda)$}&\multicolumn{1}{c}{{\%}err} & \multicolumn{1}{c}{$I(\lambda)$} 	& \multicolumn{1}{c}{{\%}err} \\ \hline
\endhead
3587	& \ion{He}{1}		& 0.214		& 0.133	&13	&0.217	&15	&0.118	&26	&0.160 	&18\\
3614	& \ion{He}{1}		& 0.209		& 0.261 &9	&\nodata&\nodata&0.377	&15	&0.259 	&14\\
3634	& \ion{He}{1}		& 0.206		& 0.255 &9	&0.262	&13	&\nodata&\nodata&0.282	&13\\
3676	& H 22			& 0.199		& 0.448 &7	&\nodata&\nodata&\nodata&\nodata&\nodata&\nodata\\
3679	& H 21			& 0.198		& 0.530 &6	&0.517	&9	&0.491	&13	&0.479	&10\\
3683	& H 20 			& 0.198		& 0.556	&6	&0.530	&9	&0.564	&12	&0.520	&10\\
3687	& H 19			& 0.197		& 0.662	&6	&0.621	&9	&0.666	&11	&0.616	&9\\
3692	& H 18			& 0.196		& 0.815	&5	&0.751	&8	&0.805	&10	&0.763	&8\\
3697	& H 17			& 0.195		& 0.948	&5	&0.901	&7	&0.886	&10	&0.900	&7\\
3704	& H 16			& 0.194		& 1.539	&4	&1.518	&6	&1.522	&7	&1.524	&6\\
3712	& H 15			& 0.192		& 1.402	&4	&1.363	&6	&1.413	&8	&1.370	&6\\
3722	& H 14 + [\ion{S}{2}] 	& 0.190 		& 0.628	&6	&1.619	&5	&1.823	&7	&1.470	&6\\
3726	& [\ion{O}{2}] 		& 0.190		& 85.732&1	&60.557	&1	&66.532	&1	&63.628	&1\\ 
3729	& [\ion{O}{2}]		& 0.189		& 52.546&1	&29.413	&2	&29.748	&2	&28.685	&2\\ 
3734	& H 13			& 0.189 		& 2.348 &3	&2.119	&5	&2.092	&7	&2.229	&5\\
3750	& H 12			& 0.186 	& 3.086	&3	&3.118	&4	&3.093	&6	&3.070	&4\\
3770	& H 11			& 0.182		& 3.952 &3	&3.958	&4	&3.923	&5	&3.917	&4\\
3798	& H 10			& 0.177		& 5.218 &2	&5.236	&3	&5.188	&4	&5.156	&3\\
3820	& \ion{He}{1}		& 0.174		&  1.066&5	&1.097	&7	&1.073	&9	&1.095	&7\\
3836	& H 9			& 0.171		& 7.329 &2	&7.208	&3	&7.303	&4	&7.272	&3\\
3856	& \ion{Si}{2}		& 0.167		& \nodata &\nodata&0.283&13	&0.297	&16	&0.185	&16\\
3863	& \ion{Si}{2}		& 0.166		& \nodata &\nodata&\nodata&	&0.175	&21	&\nodata	&\nodata\\
3869	& [\ion{Ne}{3}]		& 0.165		& 9.656	&2	&10.583	&2	&8.960	&3	&13.841	&2\\
3889	& H 8 + \ion{He}{1} 	& 0.162		&  17.338&1	&15.693	&2	&15.250	&3	&16.300	&2\\
3919	& \ion{C}{2}?		& 0.156		& 0.115 & 13	&0.162	&17	&0.186	&20	&0.146	&18\\
3927	& \ion{He}{1}		& 0.155		& 0.096	&15	&0.122	&19	&0.107	&27	&0.086	&24\\	
3933	&\ion{O}{1}		& 0.154		&\nodata	&\nodata	&0.099	&21	&0.249	&18	&\nodata	&\nodata\\
3970	& [Ne III] + H 7 	& 0.148		& 20.435&1	&20.452	&2	&19.940	&2	&21.698	&2\\
3993	& [\ion{Ni}{2}]		& 0.144		&\nodata&\nodata&\nodata&\nodata&0.063	&35	&\nodata	&\nodata\\
4009	& \ion{He}{1}		& 0.141		& 0.288 &10	&0.492	&11	&0.661	&12	&0.288	&16\\
4026	& \ion{He}{1}		& 0.138		& 1.974 &3	&2.119	&5	&2.038	&6	&2.091	&5\\
4069	& [\ion{S}{2}]		& 0.130		& 1.255	 &4	&2.454	&4	&4.200	&4	&1.642	&5\\
4076	& [\ion{S}{2}]		& 0.129		& 0.445	 &7	&0.928	&7	&1.548	&7	&0.608	&9\\
4102	& H$\delta$		& 0.125		& 26.277 &1	&26.249	&2	&25.970	&2	&26.779	&2\\
4114	& [\ion{Fe}{2}]		&0.122		&\nodata	&\nodata	&\nodata&	&0.119	&25	&\nodata	&\nodata\\
4121	& \ion{He}{1}		& 0.121		& 0.180	&11	&0.226	&14	&0.220	&19	&0.209	&15\\
4131	& \ion{O}{2}		& 0.119		&\nodata & \nodata &\nodata&	&\nodata&\nodata&0.037	&36  \\
4144	& \ion{He}{1}		& 0.117		& 0.238	& 9	&0.259	&13	&0.256	&17	&0.269	&13\\
4155	& \ion{O}{2} + \ion{N}{2}&0.116		&\nodata& \nodata &\nodata&\nodata&0.058&36	&0.051	&30 \\
4169	& \ion{O}{2}		& 0.113		& 0.039	& 22	&\nodata&\nodata&0.027	&53	&0.043	&33\\
4178	& [\ion{Fe}{2}]		&0.112		&\nodata&\nodata&\nodata&\nodata&0.076	&31	&\nodata	&\nodata  \\
4244	& [\ion{Fe}{2}]+[\ion{Fe}{3}]& 0.100		& 0.327	&8	&0.133	&18	&0.432	&13	&\nodata	&\nodata\\
4249	& [\ion{Ni}{2}] + [\ion{Fe}{2}]	& 0.099		&\nodata&\nodata&0.232	&14	&0.223	&18	&0.183	&16\\
4267	& \ion{C}{2}			& 0.096		&0.234 	&9	&0.209	&14	&0.213	&19	&0.216	&15\\
4277	& [\ion{Fe}{2}]		& 0.095		&0.045	&21	&0.064	&26	&0.215	&19	&0.041	&34\\
4287	& [\ion{Fe}{2}]		& 0.093		& 0.112	&13	&0.108	&20	&0.449	&13	&0.082	&24\\
4320	& [\ion{Fe}{2}]		&0.088		&\nodata	&\nodata	&\nodata&\nodata&0.153	&22	&\nodata	&\nodata	\\
4326	& [\ion{Ni}{2}]		& 0.086		&\nodata&\nodata&0.123	&18	&0.274	&16	&0.053	&29\\
4340	& H$\gamma$		& 0.084		& 46.866&1	&47.007	&1	&46.735	&2	&46.622	&1\\
4353	&[\ion{Fe}{2}]		&0.082		&\nodata	&\nodata	&\nodata&\nodata&0.119	&25	&\nodata	&\nodata\\
4363	& [\ion{O}{3}]		& 0.080		& 0.829	&5	&1.033	&6	&0.903	&9	&1.060	&7\\
4388	& \ion{He}{1}		& 0.076		& 0.515	&7	&0.613	&9	&0.583	&12	&0.586	&9\\
4415	& \ion{O}{2}		& 0.072		& 0.146	&11	&0.235	&13	&0.607	&11	&0.129	&19\\
4438	& \ion{He}{1}		& 0.068		& 0.051	&19	&0.059	&26	&0.050	&38	&0.065	&26\\
4452	& [\ion{Fe}{2}]		& 0.065		&\nodata&\nodata&0.058	&27	&0.157	&21	&0.032	&37\\
4458	& [\ion{Fe}{2}]		&0.062		&\nodata	&\nodata	&0.050	&29	&0.136	&23	&\nodata	&\nodata\\
4471	& \ion{He}{1}		& 0.055		& 4.350	&2	&4.708	&3	&4.494	&4	&4.594	&3\\
4515	&[\ion{Fe}{2}]		&0.046		&\nodata&\nodata&\nodata&\nodata&0.040	&42	&\nodata	&\nodata \\
4571	& \ion{Mg}{1}]		& 0.044		&\nodata&\nodata&\nodata&\nodata&0.233	&17	&\nodata	&\nodata\\
4581	&[\ion{Cr}{3}]		&0.042		&\nodata&\nodata&\nodata&\nodata&0.037	&44	&\nodata	&\nodata \\
4595	& [Co IV] ?		& 0.042		&\nodata&\nodata&0.060	&26	&0.090	&28	&0.018	&49\\ 
4607	& [\ion{Fe}{3}]		& 0.040		&0.049 	&19	&0.308	&11	&0.597	&11	&0.051	&29\\
4630	& \ion{N}{2}		& 0.036		&0.027	&26	&0.029	&38	&0.031	&48	&0.033	&36\\
4642	& \ion{O}{2} 		& 0.035		&0.109 	&13	&0.132	&17	&0.162	&21	&0.145	&17\\
4650	& \ion{O}{2}		& 0.033		&0.102 	&13	&0.143	&17	&0.134	&23	&0.137	&18\\
4658	& [\ion{Fe}{3}]		& 0.032		&0.793 	&5 	&4.896	&3	&9.533	&3	&0.711	&8\\
4665	& [\ion{Fe}{3}]		& 0.031		&\nodata&\nodata&0.198	&14	&0.444	&13	&0.018	&49\\
4701	& [\ion{Fe}{3}] 		& 0.025		&0.219 	&9	&1.716	&5	&3.311	&5	&0.220	&14\\
4711	& [\ion{Ar}{4}]+\ion{He}{1}& 0.023	&0.531 	&6	&0.570	&8	&0.546	&11	&0.645	&8\\
4728	& [\ion{Fe}{2}] 		& 0.021		&\nodata&\nodata&\nodata&\nodata&0.103	&26	&\nodata	&\nodata\\
4734	& [\ion{Fe}{3}] 	& 0.020		&0.072 	&16	&0.781	&7	&1.502	&7	&0.080	&23\\
4740	& [\ion{Ar}{4}] 	& 0.019		&0.021	&29	&\nodata&\nodata&\nodata&\nodata&\nodata	&\nodata\\
4755	& [\ion{Fe}{3}]		& 0.017		&0.142 	&11	&0.937	&7	&1.753	&6	&0.140	&17\\
4770	& [\ion{Fe}{3}]		& 0.014		&0.070 	&16	&0.600	&8	&1.165	&8	&0.071	&25\\
4779	& [\ion{Fe}{3}]		& 0.013		&0.044	&20	&0.383	&10	&0.808	&9	&0.043	&31\\
4797	& \ion{Cl}{1}		& 0.010		&0.049 	&19	&\nodata&\nodata&0.069	&32	&\nodata	&\nodata\\
4800	& \ion{O}{1}?		& 0.009		&\nodata&\nodata&0.076	&23	&\nodata&\nodata&0.056	&28\\
4815	& [\ion{Fe}{2}]		& 0.007		&0.065 	&17	&0.104	&19	&0.328	&15	&0.047	&30\\
4861	& H$\beta$		& 0.000		&98.818	&1 	&100.004&1	&99.705	&1	&98.332	&1\\
4874	&[\ion{Fe}{2}]		&0.000		&\nodata&\nodata&\nodata&\nodata&0.108	&25	&\nodata	&\nodata \\
4881	& [\ion{Fe}{3}]		& -0.001	&0.276 	&8	&2.625	&4	&5.003	&4	&0.292	&12\\
4890	& [\ion{Fe}{2}]		& -0.001	&0.033 	&23	&0.070	&24	&0.229	&17	&0.022	&43\\
4895	&[\ion{Fe}{2}]+[\ion{Cr}{3}]&-0.001	&0.047	&19	&\nodata&\nodata&\nodata&\nodata&\nodata	&\nodata\\
4905	& [\ion{Fe}{2}]		& -0.001	&0.021	&29	&0.078	&22	&0.145	&22	&0.029	&38\\
4922	& \ion{He}{1}		& -0.002	&1.205 	&4	&1.325	&6	&1.307	&8	&1.272	&6\\
4931	& [\ion{Fe}{3}]		& -0.002	&0.065 	&16	&0.262	&12	&0.523	&11	&0.053	&28\\
4959	& [\ion{O}{3}]		& -0.004	&85.528 &1	&96.013	&1	&79.275	&1	&101.927&1\\
4987	& [\ion{Fe}{3}]		& -0.006	&0.091 	&14	&0.448	&9	&0.916	&9	&0.050	&29\\
5007	& [\ion{O}{3}]		& -0.007	&253.745 &1	&283.142&1	&238.115&1	&304.533&1\\
5016	& \ion{He}{1}		& -0.008	&2.360 	&3	&2.501	&4	&2.347	&6	&2.456	&4\\
5041	& \ion{Si}{2}		& -0.010	&0.104 	&13	&0.143	&16	&\nodata&\nodata&0.064	&26\\
5048	& \ion{He}{1}		& -0.010	&0.240 	&10	&0.253	&16	&0.240	&22	&0.250	&16\\
5056	& \ion{Si}{2}		& -0.011	&0.181 	&10	&0.236	&13	&0.253	&16	&0.139	&17\\
5085	& [\ion{Fe}{3}]		& -0.014	&\nodata&\nodata&\nodata&\nodata&0.288	&15	&\nodata	&\nodata\\
5112	& [\ion{Fe}{2}]		& -0.017	&0.031	&24	&0.028	&37	&0.157	&21	&\nodata	&\nodata\\
5147	& \ion{O}{2}		& -0.020	&0.044 	&20	&0.039	&31	&\nodata&\nodata&0.034	&35\\
5159	& [\ion{Fe}{2}]		& -0.022	&0.097 	&13	&0.280	&12	&1.000	&8	&0.067	&25\\
5192	& [\ion{Ar}{3}]		& -0.026	&0.046	&19	&0.062	&25	&0.042	&40	&0.056	&27\\
5198	& [\ion{N}{1}]		& -0.027	&0.518 	&6	&0.218	&13	&0.230	&17	&0.293	&12\\
5220	& [\ion{Fe}{2}]		& -0.029	&\nodata&\nodata&0.029	&36	&0.096	&26	&\nodata	&\nodata\\
5262	& [\ion{Fe}{2}]		& -0.035	&0.085 	&14	&0.120	&18	&0.457	&12	&0.046	&30\\
5270	& [\ion{Fe}{3}]		& -0.036	&0.409 	&7	&2.918	&4	&5.721	&4	&0.387	&10\\
5299	& \ion{O}{1}		& -0.040	&0.039 	&21	&\nodata&\nodata&0.070	&31	&0.027	&39\\
5334	& [\ion{Fe}{2}]		& -0.045	&\nodata&\nodata&0.067	&24	&0.243	&16	&0.018	&48\\
5376	& [\ion{Fe}{2}]		& -0.052	&\nodata&\nodata&0.029	&36	&0.164	&15	&\nodata	&\nodata\\
5412	& [\ion{Fe}{3}]		& -0.058	&0.018	&30	&0.276	&12	&0.583	&8	&0.018	&33\\
5433	& [\ion{Fe}{2}]		& -0.061	&\nodata&\nodata&\nodata&\nodata&0.077	&21	&0.017	&35\\
5455	& [\ion{Cr}{3}]		& -0.064	&\nodata&\nodata&\nodata&\nodata&0.067	&23	&0.009	&46\\
5472	& [\ion{Cr}{3}]		& -0.067	&\nodata&\nodata&\nodata&\nodata&0.090	&20	&\nodata	&\nodata\\
5485	& [\ion{Cr}{3}]		& -0.069	&\nodata&\nodata&\nodata&\nodata&0.050	&27	&\nodata	&\nodata\\
5496	& [\ion{Fe}{2}]		&-0.071		&\nodata&\nodata&\nodata&\nodata&0.036	&31	&\nodata	&\nodata\\
5507	& [\ion{Cr}{3}]		& -0.073	&\nodata&\nodata&\nodata&\nodata&0.137	&16	&\nodata	&\nodata\\
5513	& \ion{O}{1}		& -0.075	& 0.032	&23	&\nodata&\nodata&\nodata&\nodata&0.022	&30\\
5518	& [\ion{Cl}{3}]		& -0.075	& 0.466	 &6	&0.376	&10	&0.375	&10	&0.416	&7\\
5527	& [\ion{Fe}{2}]		& -0.077	&\nodata&\nodata&0.061	&25	&0.267	&11	&0.015	&37\\
5538	& [\ion{Cl}{3}]		& -0.079	&0.468 	&6	&0.527	&8	&0.534	&8	&0.547	&6\\
5552	& [\ion{Cr}{3}]		& -0.081	&\nodata &\nodata&0.115	&18	&0.260	&12	&\nodata	&\nodata\\
5555	& \ion{O}{1}		& -0.082	&0.041	&20	&\nodata&\nodata&\nodata&	&0.029	&26\\
5667	& \ion{N}{2}		& -0.103	&0.021	&28	&\nodata&\nodata&0.037	&30	&0.028	&27\\
5680	&\ion{N}{2}		&-0.105		&\nodata	&\nodata	&\nodata	&\nodata	&0.036	&31	&0.034	&24\\
5715	& [\ion{Cr}{3}]		& -0.106	&\nodata&\nodata&\nodata&\nodata&0.124	&17	&\nodata	&\nodata\\
5747	& [\ion{Fe}{2}]		& -0.111	&\nodata&\nodata&\nodata&\nodata&0.048	&26	&\nodata	&\nodata\\
5755	& [\ion{N}{2}]		& -0.120	&0.735 	&5	&0.965	&6	&1.302	&5	&0.790	&5\\
5867	&\ion{O}{1}		&-0.130		&\nodata	&\nodata&\nodata&\nodata&\nodata& \nodata&0.037	&23\\
5876	& \ion{He}{1}		& -0.144	& 12.920&1	&13.351	&2	&13.535	&2	&13.614	&2\\
5932	& \ion{N}{2}		& -0.155	&\nodata&\nodata&\nodata&\nodata&\nodata&\nodata&0.018	&33\\
5942	& \ion{N}{2}		& -0.157	&\nodata&\nodata&\nodata&\nodata&\nodata&\nodata&0.026	&27\\
5958	& \ion{Si}{2} + \ion{O}{1}& -0.160	& \nodata&\nodata&\nodata&\nodata&0.153	&15	&0.076	&16\\ 
5979	& \ion{Si}{2}		& -0.165	&\nodata&\nodata&\nodata&\nodata&0.220	&12	&0.086	&15\\
6000	& [\ion{Ni}{3}]		& -0.169	&\nodata&\nodata&\nodata&\nodata&0.122	&16	&0.013	&38\\
6046	& \ion{O}{1}		& -0.179	&\nodata&\nodata&\nodata&\nodata&0.077	&20	&0.077	&15\\
6312	& [\ion{S}{3}]		& -0.234	&1.560 	&3	&1.966	&4	&1.935	&4	&1.760	&3\\
6347	& \ion{Si}{2}		& -0.242	&\nodata &\nodata&0.251	&11	&0.300	&10	&0.132	&12\\
6371	& \ion{Si}{2}		& -0.247	&\nodata &\nodata&0.129	&16	&0.146	&14	&0.067	&16\\
6400	& [\ion{Ni}{3}]		& -0.253	&\nodata&\nodata&0.055	&24	&0.083	&19	&0.014	&35\\
6440	&[\ion{Fe}{2}]		&-0.261		&\nodata&\nodata&\nodata&\nodata&0.052	&24	&\nodata	&\nodata \\
6534	& [\ion{Ni}{3}]		& -0.281	&\nodata&\nodata&\nodata&\nodata&0.178	&13	&\nodata	&\nodata\\
6548	& [\ion{N}{2}]		& -0.284	&18.984 &1	&18.795	&2	&23.987	&1	&16.555	&1\\
6563	& H$\alpha$		& -0.287	&288.294 &1	&289.239&1	&288.392&1	&288.675&1\\
6578	& \ion{C}{2}		& -0.290	&0.337	&7	&\nodata&\nodata&\nodata&\nodata&0.281	&8\\
6583	& [\ion{N}{2}]		& -0.291	&56.261 &1	&57.606	&1	&72.984	&1	&49.607	&1\\
6669	& [\ion{Ni}{2}]		& -0.308	&\nodata&\nodata&\nodata&\nodata&0.101	&17	&0.020	&29\\
6678	& \ion{He}{1}		& -0.310	&3.375	&2	&3.658	&3	&3.516	&3	&3.590	&2\\
6716	& [\ion{S}{2}]		& -0.318	&5.154 	&2	&3.347	&3 	&4.480	&3	&2.918	&3\\
6731	& [\ion{S}{2}]		& -0.321	&7.142 	&2	&6.040	&2	&8.430	&2	&4.990	&2\\
6797	&[\ion{Ni}{3}]		&-0.334		&\nodata&\nodata&\nodata&\nodata&0.022	&36	&\nodata	&\nodata\\
6946	& [\ion{Ni}{3}]		& -0.364	&\nodata&\nodata&0.029	&32	&\nodata&\nodata&\nodata	&\nodata\\
7002	& \ion{O}{1}		& -0.375	& 0.100	&11	&\nodata&\nodata&0.077	&19	&0.063	&16\\
7065	& \ion{He}{1}		& -0.388	& 4.297	&2	&4.802	&3	&4.582	&3	&5.536	&2\\
7110	& \ion{O}{2}?		& -0.396	&\nodata&\nodata&\nodata&\nodata&0.046	&24	&0.044	&19\\
7136	& [\ion{Ar}{3}]		& -0.401	&11.611 &1	&13.700	&2	&12.382	&2	&13.967	&1\\
7155	& [\ion{Fe}{2}]		& -0.405	&0.059 	&15	&0.304	&10	&1.436	&4	&0.050	&17\\
7161	& \ion{He}{1}		& -0.410	&\nodata&\nodata&\nodata&\nodata&\nodata&\nodata&0.018	&29\\
7231	& \ion{C}{2}		& -0.419	&0.063	&14	&0.093	&17	&0.078	&18	&0.078	&14\\
7236	& \ion{C}{2}		& -0.421	& 0.157	&9	&0.134	&14	&0.097	&16	&0.148	&10\\
7254	& \ion{O}{1}		& -0.424	& \nodata	&\nodata	&\nodata&\nodata&0.111	&15	&0.081	&14\\
7281	& \ion{He}{1}		& -0.429	& 0.554	&5	&0.636	&7	&0.607	&7	&0.621	&5\\
7291	& [\ion{Ca}{2}]		& -0.431	&\nodata&\nodata&0.138	&14	&0.572	&7	&\nodata	&\nodata\\
7298	& \ion{He}{1}		& -0.432	&0.032 	&20	&\nodata&\nodata&\nodata&\nodata&0.035	&21\\
7320	& [\ion{O}{2}]		& -0.436	&3.673 	&2	&6.784	&2	&9.330	&2	&5.244	&2\\
7330	& [\ion{O}{2}]		& -0.438	&2.977 	&2	&5.694	&2	&7.656	&2	&4.305	&2\\
7341	&\ion{Ca}{1}		&-0.440		&\nodata&\nodata&\nodata&\nodata&\nodata&\nodata&0.042	&19\\
7370	&\ion{Ca}{1}		&-.445		&\nodata&\nodata&\nodata&\nodata&\nodata&\nodata&0.025	&24\\
7378	& [\ion{Ni}{2}]		& -0.447	&0.080	&12	&\nodata&\nodata&1.297	&5	&0.067	&15\\
7388	& [\ion{Fe}{2}]		& -0.449	&\nodata&\nodata&0.079	&18	&0.285	&10	&\nodata	&\nodata\\
7402	&\ion{Ca}{2}		&-0.450		&\nodata&\nodata&\nodata&\nodata&\nodata&\nodata&0.018	&29\\
7412	& [\ion{Ni}{2}]		& -0.453	&\nodata&\nodata&\nodata&\nodata&0.137	&14	&0.019	&28\\
7424	&\ion{Ca}{1}		&-0.456		&\nodata&\nodata&\nodata&\nodata&\nodata&\nodata&0.007	&47\\
7443	&\ion{S}{1}		&-0.459		&\nodata&\nodata&\nodata&\nodata&\nodata&\nodata&0.017	&29\\
7453	& [\ion{Fe}{2}]		& -0.460	& \nodata&\nodata&0.091	&17	&0.442	&8	&0.015	&32\\
\midrule
\multicolumn{3}{c}{$I($H$\beta)$\tabnotemark{b}}	& 	\multicolumn{2}{c}{8.96E-012}	& \multicolumn{2}{c}{1.55E-012}	& \multicolumn{2}{c}{1.49E-012}		& \multicolumn{2}{c}{ 1.80E-011} \\
\multicolumn{3}{c}{$C$(H$\beta$)} & \multicolumn{2}{c}{0.35}	&\multicolumn{2}{c}{0.36}	& \multicolumn{2}{c}{0.37} & \multicolumn{2}{c}{0.35}\\
\multicolumn{3}{c}{EW$_\mathrm{abs}(\mathrm{H}\beta)$}& \multicolumn{2}{c}{3.8} & \multicolumn{2}{c}{6.6}	& \multicolumn{2}{c}{7.0} & \multicolumn{2}{c}{4.4} \\
\bottomrule
\tabnotetext{a}{\scriptsize Emission lines corrected for reddening, underlying absorption and normalized with respect to the entire Balmer decrement.}
\tabnotetext{b}{\scriptsize Dereddened flux for H$\beta$ in units of erg cm$^{-2}$ s$^{-1}$. The North and South zones are the sum of four 50-pixel long spectra from the Orion Nebula. The weakly and strongly shocked zones are the sum of four and three 3 pixel long spectra respectively covering HH202-S (see text).}
\end{longtable}
\end{changemargin}

\normalsize

\subsection{Chemical composition}

Ionic abundances were derived for O$^{+}$, O$^{2+}$, N$^{+}$, Ne$^{2+}$, S$^{+}$, S$^{2+}$, Cl$^{2+}$, Ar$^{2+}$, Fe$^{+}$, Fe$^{2+}$, and Ni$^{2+}$, from CELs using PyNeb. Just like in the previous section, oxygen abundances were also computed using recombination lines from those of multiplet 1. The Fe$^{+}$ abundance was estimated using only [\ion{Fe}{2}] $\lambda$7155 as it is the only line available in our observed range not affected by fluorescence; meanwhile, the Fe$^{2+}$ abundance was determined using the emission of [\ion{Fe}{3}] $\lambda$4734, $\lambda$4755 and $\lambda$4881, since these lines are not contaminated by emission of other ions. 

The He$^{+}$ abundance was calculated from recombination lines using HELIO13, a software package described in \citet{2012ApJ...753...39P} that uses a maximum likelihood method to perform a simultaneous fitting of $n_{\mathrm{e}}$, $\tau_{3889}$, the He$^{+}$ abundance, $t^{2}$, and $T_{0}$. We present the final adopted value of $t^{2}$ for each region in Table \ref{table.t2}. In Table \ref{table.ionabundances}, we present the ionic abundances assuming both homogeneous and inhomogeneous temperature distributions: as mentioned in section 3 we prefer abundance determinations with $t^{2} \ne $ 0.00 because the shocked region obviously has an inhomogeneous temperature distribution, also O$^{2+}_{\mathrm{CEL}}$ abundances agree better with the RL O$^{2+}_{\mathrm{RL}}$ abundances when $t^{2} \ne $ 0.00 is used.

\begin{table}\centering
\setlength{\tabnotewidth}{\columnwidth}
  \tablecols{6}
\caption{Physical conditions from combined spectra}
\label{table.temden}
\begin{tabular}{c l l r r r}
\toprule
\multicolumn{2}{c}{Diagnostic} & North Zone & South Zone & WS Zone & SS Zone \\
\midrule
$T_{\mathrm{e}}$ (K) 		& [\ion{O}{3}] 	& 8210 $\pm$ 120  & 8320 $\pm$ 170 	   & 8410 $\pm$ 160  	    & 8490 $\pm$ 220  \\
				& [\ion{Ar}{3}] & 8200 $\pm$ 440  & 8220 $\pm$ 650 	   & 8530 $\pm$ 670  	    & 7830 $\pm$ 890 \\
				& [\ion{N}{2}]	& 9370 $\pm$ 200  & 9800 $\pm$ 300 	   & 10060 $\pm$ 300        & 10125 $\pm$ 240 \\
				& [\ion{S}{2}]	& 10180 $\pm$ 400  & 11680 $\pm$ 800 	   & 19920 $\pm^{2500}_{1500}$& 17710 $\pm$ 1000 \\
				& [\ion{O}{2}]	& 13120 $\pm$ 600  & 14700 $\pm$ 900 	   & 23150 $\pm$ 1500 	    & 21400 $\pm$ 2200 \\
				& Adopted HI	& 8210 $\pm$ 150  &  8320 $\pm$ 200	   & 	8410 $\pm$ 200      &	8490 $\pm$ 300 \\
				& Adopted LI	& 9500 $\pm$ 250  & 9780 $\pm$	300	   &	10230 $\pm$ 300     & 10260 $\pm$ 350 \\
$n_{\mathrm{e}}$ (cm$^{-3}$)	& [\ion{Cl}{3}] & 2500 $\pm$ 1100 & 5980$^{+2700}_{-1900}$ & 7200$^{+5100}_{-3000}$ & 7550$^{+5300}_{-3000}$ \\
				& [\ion{O}{2}]	& 1860 $\pm$ 110  & 5720 $\pm$ 800 	   & 4230 $\pm$ 500	    & 6300 $\pm$ 900 \\
				& [\ion{S}{2}]  & 1530 $\pm$ 200  & 3500 $\pm$ 600	   & 4590 $\pm$ 1100	    & 5800 $\pm$ 1500 \\
				& Adopted HI	& 2500 $\pm$ 1000  & 5980 $\pm$ 2000	   &	7200 $\pm$ 3000	    &	7550 $\pm$ 3000		\\
				& Adopted LI	& 1800 $\pm$ 200  & 4750 $\pm$ 500	   &	4250 $\pm$ 600	    &	6100 $\pm$ 900		\\ 
\bottomrule 
\end{tabular}
\end{table}

In order to calculate total abundances, we have to consider the contribution from unseen ions; this is done assuming a series of ionization correction factors (ICFs) from different sources. We do not expect this to be the case of oxygen, whose total abundance of which is simply the sum of O$^{+}$ and O$^{2+}$.

For nitrogen we have used the classic ICF, to account for the presence of N$^{2+}$:
\begin{equation}
\frac{\mathrm{N}}{\mathrm{H}} = \left[ \frac{\mathrm{O}^{+} + \mathrm{O}^{2+}}{\mathrm{O}^{+}} \right]_{\mathrm{CEL}} \times \frac{\mathrm{N}^{+}}{\mathrm{H}^{+}} = \mathrm{ICF}(\mathrm{N}^{2+}) \times \frac{\mathrm{N}^{+}}{\mathrm{H}^{+}}.
\end{equation}

The total neon abundance has a contribution from Ne$^{2+}$, which we have taken in consideration using the ICF from \citet{1969BOTT....5....3P}:
\begin{equation}
\frac{\mathrm{Ne}}{\mathrm{H}} = \left[ \frac{\mathrm{O}^{+} + \mathrm{O}^{2+}}{\mathrm{O}^{2+}} \right]_{\mathrm{CEL}} \times \frac{ \mathrm{Ne}^{2+}}{\mathrm{H}^{+}} = \mathrm{ICF}(\mathrm{Ne}^{2+}) \times \frac{\mathrm{Ne}^{2+}}{\mathrm{H}^{+}}.
\end{equation}

Besides S$^{+}$ and S$^{2+}$, it is known that S$^{3+}$ must be present in \ion{H}{2} regions from the work of \citet{1978A&A....66..257S}:
\begin{equation}
\frac{\mathrm{S}}{\mathrm{H}} = \left[ 1 - \left[ \frac{\mathrm{O}^{+}}{\mathrm{O}^{+} + \mathrm{O}^{2+}} \right]^{3}_{\mathrm{CEL}}  \right]^{-1/3} \times \frac{\mathrm{S}^{+} + \mathrm{S}^{2+}}{\mathrm{H}^{+}} = \mathrm{ICF}(\mathrm{S}^{+} + \mathrm{S}^{2+}) \times \frac{\mathrm{S}^{+} + \mathrm{S}^{2+}}{\mathrm{H}^{+}}.
\end{equation}

Helium has to be corrected for the presence He$^{0}$, we have done this using the ICF derived by \citet{1992RMxAA..24..155P}:
\begin{equation}
\frac{\mathrm{He}}{\mathrm{H}} =   \left[1 + \frac{\mathrm{S}^{+}}{\mathrm{S} - \mathrm{S}^{+}} \right] \times \frac{\mathrm{He}^{+}}{\mathrm{H}^{+}} = \mathrm{ICF}(\mathrm{He}^{+}) \times \frac{\mathrm{He}^{+}}{\mathrm{H}^{+}},
\end{equation}
however, as \citet{2014MNRAS.440..536D} point out, this result has to be taken with reservation, since the population of helium ions depends on the effective temperature of the ionizing stars, and that of sulfur on the ionization parameter. 

We have observed [\ion{Cl}{3}] emission lines in our spectra, however Cl$^{+}$ and possibly Cl$^{3+}$ also contribute to the total chlorine abundance. \citet{2014MNRAS.440..536D} propose an ICF which, although intended for use in planetary nebulae, can be used in this case as it depends on the ionic fraction of oxygen and the observed abundance of Cl$^{2+}$:
\begin{eqnarray}
\frac{\mathrm{Cl}}{\mathrm{H}} &=& \left(4.1620 - 4.1622 \left[\frac{\mathrm{O}^{2+}}{\mathrm{O}^{+} + \mathrm{O}^{2+}} \right]^{0.21}\right)^{0.75} \times \frac{\mathrm{Cl}^{2+}}{\mathrm{O}^{+}} \times \frac{\mathrm{O}^{+} + \mathrm{O}^{2+}}{\mathrm{H}^{+}} \nonumber \\ 
&=& \mathrm{ICF}\left( \frac{\mathrm{Cl}^{2+}}{\mathrm{O}^{+}} \right) \times \frac{\mathrm{Cl}^{2+}}{\mathrm{O}^{+}} \times \frac{\mathrm{O}^{+} + \mathrm{O}^{2+}}{\mathrm{H}^{+}};
\end{eqnarray}
this is valid when the ionic fraction of oxygen ---the term in square brackets--- takes a value between 0.02 and 0.95.

It is known that Ar$^{+}$ can contribute a significant fraction to the total abundance.  For this work, we have employed the ICF obtained by \citet{2014MNRAS.440..536D} from Cloudy photoionization models, which depends only on Ar$^{2+}$ lines:

\begin{equation}
\frac{\mathrm{Ar}}{\mathrm{H}} = 10^{\left( \frac{0.3 \omega}{0.4 - 0.3 \omega} - 0.05 \right)} \times \frac{\mathrm{Ar}^{2+}}{\mathrm{H}^{+}} \nonumber \\
= \mathrm{ICF}(\mathrm{Ar}^{2+}) \times \frac{\mathrm{Ar}^{2+}}{\mathrm{H}^{+}}
\end{equation}
with $\omega = \mathrm{O}^{2+} / (\mathrm{O}^{+} + \mathrm{O}^{2+})$.

Uncertainties in atomic data affect some ions more than others; given the complex structure of Fe$^{+}$ and Fe$^{2+}$, the atomic data currently available does not yet represent a complete picture of this element; furthermore, it is known that many [\ion{Fe}{2}] lines are affected by fluorescence. For computing the total iron abundance, \citet{2005ApJ...626..900R} propose two ICFs based on observations and photoionization models which require only [\ion{Fe}{III}] lines. We have decided to use their observational ICF since the one from the models produces results for total iron abundance that would imply a complete destruction of interstellar dust grains (see section \ref{subsec.dustdestruction}), something that is not expected in these observations since we have substantial amounts of unshocked gas in front of and behind the shocked region; also this shock is not fast enough to be expected to destroy all the dust grains it encounters. Thus
\begin{equation}
\frac{\mathrm{Fe}}{\mathrm{H}} = 1.1 \left( \frac{\mathrm{O}^{+}}{ \mathrm{O}^{++}} \right)^{0.58} \times \frac{\mathrm{Fe}^{++}}{\mathrm{O}^{+}} \times \frac{\mathrm{O}}{\mathrm{H}}.
\end{equation}

Nickel poses similar problems to iron in that [\ion{Ni}{II}] lines may be affected by fluorescence. Until recently, most studies have used an ICF for Nickel that is based on the similarity of the ionization potentials of Fe$^{+}$ and Ni$^{+}$. Based on multiple observational data and photoionization models, \citet{2016MNRAS.456.3855D} have derived two ICFs for Ni that require only [\ion{Ni}{3}] lines.  From Equation 6 of their paper ---applicable when He$^{2+}$ is not present--- we have that
\begin{eqnarray}
\frac{\mathrm{Ni}}{\mathrm{H}} &=&  \left(1.1 - 0.9\frac{\mathrm{O}^{2+}}{\mathrm{O}^{+} + \mathrm{O}^{2+}}\right) \times \frac{\mathrm{Ni}^{2+}}{\mathrm{O}^{+}} \times (\frac{\mathrm{O}^{+} + \mathrm{O}^{2+}}{\mathrm{H^{+}}}) \\
&=& \mathrm{ICF}(\frac{\mathrm{Ni}^{2+}}{\mathrm{O}^{+}})  \times \frac{\mathrm{Ni}^{2+}}{\mathrm{O}^{+}} \times \left(\frac{\mathrm{O}^{+} + \mathrm{O}^{2+}}{\mathrm{H^{+}}} \right).
\end{eqnarray}
We have used [\ion{Ni}{3}] $\lambda$4326, $\lambda$6000, and $\lambda$6401 to calculate the total abundance since these lines are not contaminated by others in our spectra.

Total abundances are presented in Table \ref{table.totalabundances} considering a both a homogeneous ($t^{2} = 0.00$) and an inhomogeneous temperature ($\protect t^{2} \ne 0.00$).

\begin{table}
\caption{\lowercase{$t^{2}$} values}
\label{table.t2}
\begin{tabular}{c c c c}
\toprule
North Zone & South Zone & WS Zone & SS Zone \\
\midrule
0.014 $\pm$ 0.005 & 0.022 $\pm$ 0.004 & 0.024 $\pm$ 0.006 & 0.039 $\pm$ 0.006 \\
\bottomrule
\end{tabular}
\end{table}

\section{Discussion}\label{sec.discussion}

The results obtained here for $T_{e}$ and $N_{e}$ for the North and South zones agree with previous determinations by \citet{2004MNRAS.355..229E} and \citet{2003MNRAS.340..362R}. For the shocked zones we must compare our results directly with those of \citet{2009MNRAS.395..855M} as it is the only other analysis of HH 202 available in the literature; while our results are in considerable agreement for the unshocked zones, at the apex of the shock the authors of that work adopt a considerably higher density  (17 430 $\pm$ 2360 cm$^{-3}$), and use the same value as representative of both high and low ionization zones. This may indicate that the volume of gas analyzed in our work is different from that of \citet{2009MNRAS.395..855M}; also the volume of gas we examined contains both shocked and unshocked components.

As we can see in Tables \ref{table.ionabundances} and \ref{table.totalabundances}, O$^{2+}$ and O abundances determined from CELs and RLs are irreconcilable at the strongly shocked zone unless we consider the presence of temperature fluctuations. In their study of HH 202--S, \citet{2009MNRAS.395..855M} reported values for $t^{2} = 0.049$ and $t^{2} = 0.050$ at the center of the shock, which imply a greater abundance for O$_{\mathrm{CEL}} = 8.76 \pm 0.06$ that is not compatible with their estimate for O$_{\mathrm{RL}} = 8.65 \pm 0.05$. As noted in that paper, this may indicate that the t$^{2}$ paradigm is not applicable in the case of a purely shocked volume of gas.

Just as in our analysis from section 4, we find that the Abundance Discrepancy Factor (ADF) associated to O$^{2+}$ and O is greater at the apex of the shock. This connection between Herbig-Haro objects and the ADF had been reported previously by \citet{2008ApJ...675..389M} who found several above-average increases in the ADF associated with Herbig-Haro objects 202, 203, and 204 in the Orion Nebula; however, the cause behind these high ADF values ---be it temperature fluctuations, or any other mechanism--- remained uncertain. Thanks to the quality of our observations we can determine that, at the Strongly Shocked Zone, the mean squared temperature fluctuations show a peak value of $t^{2} = 0.039 \pm 0.006$ which, as can be seen in Tables \ref{table.ionabundances} and \ref{table.totalabundances}, reconciles the ionic O$^{2+}$ abundance and the total oxygen abundance determined from CELs and RLs in all of the observed zones. This result and the behavior observed in Figure \ref{fig.oxigeno_total} appear to indicate that the $t^{2}$ parameter is intrinsically linked to shocks; this suggests that shocks embedded in the structure of the nebulae may be responsible for an important fraction of the observed $t^{2}$ parameter in \ion{H}{2} regions, as well as in the observed ADF. Clearly, a similar analysis to the one performed here on other spatially resolved interstellar shocks would help to elaborate upon this possible connection.

\subsection{Dust destruction}\label{subsec.dustdestruction}

As Table \ref{longtable.emission_lines} shows, emission lines of refractory elements such as Fe and Ni are much brighter in the weakly and strongly shocked zones. Iron is excellent for studying dust destruction in this case since it is known that about 90\% of it is depleted in dust grains \citep{2005ApJ...626..900R, 2010ApJ...724..791P},

The iron and oxygen ratio can be used as an indicator of the degree of dust destruction by comparing its value in the center of HH 202-S with the surrounding gas. The total abundance of iron, however, depends on the ICF used to calculate it. \citet{2005ApJ...626..900R} derived two ICFs for this purpose based on observations and photoionization models. Recent works \citep{2004MNRAS.355..229E, 2009MNRAS.395..855M, 2016MNRAS.456.3855D} use the ICF from photoionization models. We have calculated the iron abundance and the amount of dust destruction using both: assuming thermal inhomogeneities, and comparing our value with the solar one, (Fe/O)$_{\sun}$ = -1.22 \citep{2015A&A...573A..27G,2009ARA&A..47..481A} ) we find an increase in 1.15 dex over the average iron abundance of the unshocked zones using the observational ICF, implying 57 $\pm$ 10\% of the dust is destroyed; on the other hand, the ICF derived from photionization models delivers a value of 90\% of dust destruction. The latter value is only reached by the shock of an expanding supernova, and seems extreme for a Herbig-Haro object, especially if we take into account the fact that the area we are observing includes both shocked and unshocked material. Given these results, we favored the observational ICF by \citet{2005ApJ...626..900R} and its implications.

We can analyze the amount of nickel released by the shock as well. This element is not as abundant as iron, magnesium or silicon, and it is not expected to be mixed solely with oxygen in dust grains. \citet{2015A&A...573A..26S} have derived a value for Ni$_{\sun}$ = 6.20 $\pm$ 0.04. From our determinations (using atomic data from \citet{2001bautista}), we find that  25 $\pm$ 10 \% of Ni is released by the shockwave. This suggests that the shock is not as efficient in incorporating nickel to the gas phase as iron. A deeper discussion on this subject can be found in \citet{2016MNRAS.456.3855D}.

In \ion{H}{2} regions, it is expected that iron and oxygen are found predominantly in compounds such as ferrous oxide (FeO), therefore we have assumed that dust O and Fe are destroyed in the same fraction. Considering our average abundance for North and South zones and the value at the strongly shocked zone we can extrapolate to a total destruction by taking the solar value of (Fe/O)$_{\sun}$ = -1.22 \citep{2015A&A...573A..27G,2009ARA&A..47..481A}; with these considerations we find that the O depletion factor of the ambient gas to be -0.12 $\pm$ 0.04. This represents an improvement over the result by \citet{2009MNRAS.395..855M} who report a value of -0.11$^{+0.11}_{0.14}$, using the same method, albeit observing a smaller section of HH 202-S. 

There are two other methods that can be used to estimate the amount of depletion of oxygen. The first one consists in comparing the gaseous oxygen  abundance to the oxygen abundance in the stars of the Orion Nebula.  The oxygen abundance from B-type stars of the Ori-OB1 association has been measured to be 8.74 $\pm$ 0.04 \citep{2011A&A...526A..48S}. With this reference value and our O$_{\mathrm{RL}}$ determination we find a depletion factor of -0.18 $\pm$ 0.05 dex. Using the same method, \citet{2009MNRAS.395..855M} estimate a depletion factor of -0.17 $\pm$ 0.06. 

The last method comes from the fact that dust grains contain molecules formed from Mg, Si, Fe, and O such as olivine (Mg, Fe)$_{2}$SiO$_{4}$ and pyroxene (Mg, Fe)SiO$_{3}$. The depletion factor can be estimated then from the abundances of said elements in the gas. From these assumptions, the accepted value for the depletion factor in the Orion Nebula has been measured to be -0.10 $\pm$ 0.04. The results for the depletion factors obtained through different methods are summarized in Table \ref{table.depletions}.

First we must notice that our value for the depletion factor agrees excellently with those from the other two methods, thanks to the quality of our observations and data reduction. We have calculated the weighted average of the three methods using our results and those of \citet{2009MNRAS.395..855M} from the previous paragraphs obtaining a depletion factor of -0.126 $\pm$ 0.024. 

\begin{center}
\begin{table}
\caption{Oxygen depletion factors}
\label{table.depletions}
\begin{tabular}{c c c}
\toprule
Method & Value & Reference \\
\midrule
Fe/O ratio & -0.12 $\pm$ 0.04 & This work \\
			 & -0.11$^{+0.11}_{-0.14}$ & \citet{2009MNRAS.395..855M} \\
\midrule 
Comparison with Orion stars & -0.18 $\pm$ 0.05  & This work \\
			& -0.17 $\pm$ 0.06 &\citet{2009MNRAS.395..855M} \\
\midrule
Molecular composition & -0.10 $\pm$ 0.04 & \citet{1998MNRAS.295..401E, 2009MNRAS.395..855M} \\
\bottomrule
\end{tabular}
\end{table}
\end{center}

\section{Conclusions}\label{sec.conclusions}

We have performed a long-slit spectroscopic analysis of Herbig-Haro 202 using the FORS 1 spectrograph of the VLT. We have determined the spatial variations of temperature and density across the Orion Nebula and compared these to the shock. We have shown that oxygen (O) abundances determined from collisionally excited lines and recombination lines are irreconcilable at the center of the shock unless we consider the existence of thermal inhomogeneities along the line of sight. The Abundance Discrepancy Factor associated to O$^{2+}$ and O is greater at the shock, coinciding with the peak of the $t^{2}$ parameter; this fact suggests that interstellar shocks may contribute an important fraction to the $t^{2}$ parameter. Iron (Fe) abundance also shows a peak at the center of the shock, an effect that we attribute to dust destruction by the gas flow, which releases iron into the gas phase.

Spectra from four different zones of the Orion Nebula were combined to increase the signal to noise ratio. These regions represent the center of the shock and the undisturbed gas. We identified a total of 169 different emission lines, including 159 in the strongly shocked zone, that we used to derive physical conditions with high precision.

Chemical abundances for He, O, N, Ar, Cl, Ne, S, Fe and Ni were calculated assuming both homogeneous temperature and thermal inhomogeneities. We showed that O abundances from collisionally excited lines and recombination lines can be made to agree by incorporating the $t^{2}$ parameter proposed by \citet{1967ApJ...150..825P}. Also, we have reproduced the results obtained by \citet{2009MNRAS.395..855M}, complementing that work by providing a spatial analysis of the physical conditions and oxygen abundance across HH 202 and the surrounding gas; we have also reduced the uncertainties associated with some determinations, notably the O$_{\mathrm{CEL}}$ and He$^{+}$ abundances. 

Using Fe/O as an indicator, we have shown that dust destruction is taking place at the apex of HH 202,  which amounts to 57 $\pm$ 10 \%. Comparing the abundance of Ni in the static gas with the Strongly Shocked zones we have found that 25 \% of Ni is released from dust by the gas flow, suggesting that the shock is not as efficient in incorporating Ni to the ambient gas. 

Comparing the total oxygen abundance at the center of the shock with the ambient gas, and taking the solar value as reference, we found the depletion factor of oxygen to be -0.12 $\pm$ 0.04 dex. This result is a significant improvement over previous individual determinations. We also compared the total oxygen abundance with respect to the abundance in the stars of the Orion Nebula, finding a depletion factor or -0.18 $\pm$ 0.05 dex

Finally, we averaged our results with those obtained by \citet{2009MNRAS.395..855M} using the same methods, obtaining a depletion factor for oxygen of -0.126 $\pm$ 0.024.

The authors are very grateful to the organizers of the NEBULATOM 2 school: C. Morisset, G. Stasi\'nska and C. Mendoza, for their instruction in the use of PyNeb and the newest atomic data. We express our gratitude  to M. Peimbert for his support with the observations. We also thank and anonymous referee for his helpful comments and suggestions.This work was supported by Mexican CONACyT program 000205 and PAPIIT IN 109716.
\begin{landscape}

\begin{changemargin}{-2cm}{-1cm}
\begin{table}[t]\centering
\scriptsize
\caption{Ionic abundances}
\label{table.ionabundances}
\begin{tabular}{l c c c c c c c c}
\toprule
Ion 	& \multicolumn{2}{c}{North zone} & \multicolumn{2}{c}{South zone} & \multicolumn{2}{c}{WS zone} & \multicolumn{2}{c}{SS zone} \\
\midrule
	& $t^{2}=0.00$	&	$t^{2}=0.014$ & $t^{2}=0.00$	&$t^{2}=0.022$ & $t^{2}=0.00$	& $t^{2}=0.024$ & $t^{2}=0.00$	&$t^{2}= 0.039 $\\	
\cmidrule(l){2-3}
\cmidrule(l){4-5}
\cmidrule(l){6-7}
\cmidrule(l){8-9}
Ar$^{2+}$ & 6.25 $\pm$ 0.02 & 6.34 $\pm$ 0.03	& 6.31 $\pm$ 0.03 & 6.45 $\pm$ 0.03 & 6.30 $\pm$ 0.03 & 6.44 $\pm$ 0.04& 6.23 $\pm$ 0.04 & 6.48 $\pm$ 0.04\\
Cl$^{2+}$ & 5.08 $\pm$ 0.04 & 	5.19 $\pm$ 0.05		& 5.11 $\pm$ 0.03 &  5.27 $\pm$ 0.06	    & 5.08 $\pm$ 0.03 & 5.25 $\pm$ 0.06	       & 5.06 $\pm$ 0.03 & 5.35 $\pm$ 0.07\\
Fe$^{+}$  & 4.72:            & 4.77: 		& 4.59:	      	  &  4.66: 	    & 5.34:	      & 5.41: 	       & 6.00: 		 & 6.12:\\
Fe$^{2+}$ & 5.60 $\pm$ 0.04 & 5.65 $\pm$ 0.05	& 5.52 $\pm$ 0.06 & 5.60 $\pm$ 0.06 &6.38 $\pm$ 0.03 & 6.46 $\pm$ 0.03 & 6.64 $\pm$ 0.04 & 6.77 $\pm$ 0.05\\
N$^{+}$	  & 7.13 $\pm$ 0.03 & 7.18 $\pm$ 0.03 	& 7.05 $\pm$ 0.03 & 7.12 $\pm$ 0.04 &7.05 $\pm$ 0.03 & 7.12 $\pm$ 0.03 & 7.17 $\pm$ 0.04 & 7.29 $\pm$ 0.04\\
Ne$^{2+}$ & 7.42 $\pm$ 0.03 & 7.54 $\pm$ 0.05	&7.55 $\pm$ 0.04  & 7.73 $\pm$ 0.06 &7.41 $\pm$ 0.05 & 7.61 $\pm$ 0.05 & 7.31 $\pm$ 0.06 & 7.63 $\pm$ 0.07 \\ 
Ni$^{2+}$ & 	\nodata	    & \nodata 	& 4.46 $\pm$ 0.10& 4.60 $\pm$ 0.11 & 4.99 $\pm$ 0.07	& 5.13 $\pm$ 0.07 & 5.24 $\pm$ 0.06 & 5.43 $\pm$ 0.07\\
O$^{+}$& 7.91 $\pm$ 0.02& 7.97 $\pm$ 0.04 & 7.80 $\pm$ 0.03 & 7.90 $\pm$ 0.04 & 7.69 $\pm$ 0.02 & 7.79 $\pm$ 0.03 & 7.79 $\pm$ 0.03 & 7.95 $\pm$ 0.05\\
O$^{2+}$ & 8.29 $\pm$ 0.02& 8.38 $\pm$ 0.03 & 8.34 $\pm$ 0.04 & 8.49 $\pm$ 0.05 & 8.30 $\pm$ 0.03 & 8.47 $\pm$ 0.04 & 8.20 $\pm$ 0.04 & 8.50 $\pm$ 0.05 \\
O$^{2+}_{\mathrm{RL}}$ & \multicolumn{2}{c}{8.38 $\pm$ 0.04} & \multicolumn{2}{c}{8.49 $\pm$ 0.05} & \multicolumn{2}{c}{8.47 $\pm$ 0.05} & \multicolumn{2}{c}{8.51 $\pm$ 0.06} \\
S$^{+}$ & 5.70 $\pm$ 0.02   & 5.75 $\pm$ 0.03 	 &5.63 $\pm$ 0.04 & 5.70 $\pm$ 0.06	&5.62 $\pm$ 0.02	&  5.69 $\pm$ 0.03	&5.84 $\pm$ 0.02 & 5.95 $\pm$ 0.05 \\
S$^{2+}$ & 6.94 $\pm$ 0.05    & 7.06 $\pm$ 0.06	&6.96 $\pm$ 0.06 & 7.14 $\pm$ 0.07	& 6.98 $\pm$ 0.06	&7.18 $\pm$ 0.07 	&6.95 $\pm$ 0.07 & 7.27 $\pm$ 0.09\\
He$^{+}_{\mathrm{RL}}$ & 10.926 $\pm$ 0.004 & 10.922 $\pm$ 0.004  & 10.952 $\pm$ 0.006 & 10.936 $\pm$ 0.005 & 10.951 $\pm$ 0.006 & 10.940 $\pm$ 0.006 & 10.959 $\pm$ 0.007 & 10.922 $\pm$ 0.008 \\
\bottomrule
\end{tabular}
\end{table}
\end{changemargin}
\end{landscape}

\begin{landscape}
\begin{table}\centering
\scriptsize
\caption{Total abundances}
\label{table.totalabundances}
\begin{tabular}{c c c c c c c c c}
\toprule
Element	& \multicolumn{2}{c}{North zone} & \multicolumn{2}{c}{South zone} & \multicolumn{2}{c}{WS zone} & \multicolumn{2}{c}{SS zone} \\
\midrule
	& $t^{2}=0.00$	& $t^{2}=0.014$&$t^{2}=0.00$	&	$t^{2}=0.022$ & $t^{2}=0.00$	&	$t^{2}=0.024$ & $t^{2}=0.00$	&$t^{2}=0.039$\\
\cmidrule(l){2-3}
\cmidrule(l){4-5}
\cmidrule(l){6-7}
\cmidrule(l){8-9}
Ar	& 6.31$\pm^{0.2}_{0.52}$ &	6.41$\pm^{0.2}_{0.52}$	   & 6.40 $\pm^{0.2}_{0.52}$ &	6.55$\pm^{0.2}_{0.52}$	       &6.40 $\pm^{0.2}_{0.52}$&	6.56$\pm^{0.2}_{0.52}$	  & 6.28$\pm^{0.2}_{0.52}$	&	6.58$\pm^{0.2}_{0.52}$	\\
Cl	& 5.22 $\pm^{0.06}_{0.14}$& 5.32$\pm^{0.06}_{0.14}$	 & 5.26 $\pm^{0.06}_{0.14}$ & 	5.42$\pm^{0.06}_{0.14}$	       &5.24 $\pm^{0.06}_{0.14}$ & 5.42	$\pm^{0.06}_{0.14}$	&5.19$\pm^{0.06}_{0.14}$	& 5.50 $\pm^{0.06}_{0.14}$\\
Fe	& 5.95 $\pm$ 0.04& 6.00 $\pm$ 0.06 & 5.91 $\pm$ 0.06   & 5.99 $\pm$ 0.08 & 6.78 $\pm$ 0.05  & 6.87 $\pm$ 0.05 & 7.00 $\pm$ 0.06 & 7.15 $\pm$ 0.07 \\
N	& 7.67 $\pm$ 0.03& 7.73	$\pm$ 0.05 & 7.71 $\pm$ 0.02 & 7.81	$\pm$ 0.06&7.76 $\pm$ 0.03 & 7.88 $\pm$	0.05 &7.71 $\pm$ 0.02 	& 7.95 $\pm$ 0.06\\
Ne	& 7.57 $\pm$ 0.03& 7.69 $\pm$ 0.04 &  7.66 $\pm$ 0.05 & 7.83 $\pm$ 0.06   &7.51  $\pm$ 0.04 & 7.69 $\pm$ 0.05 &7.45  $\pm$ 0.07 	& 7.74 $\pm$ 0.08\\
Ni	& \nodata	& \nodata	&  4.72 $\pm$ 0.03& 4.88 $\pm$ 0.04& 5.28 $\pm$ 0.04 & 5.45 $\pm$ 0.03	& 5.45 $\pm$ 0.04& 5.69 $\pm$ 0.04\\
O$_{\mathrm{CEL}}$& 8.44 $\pm$ 0.02 & 	8.52 $\pm$ 0.02	&8.45 $\pm$ 0.03 & 8.59 $\pm$ 0.04 & 8.39 $\pm$ 0.03 & 8.56 $\pm$ 0.03 	& 8.34 $\pm$ 0.03 & 8.61 $\pm$ 0.04\\ 
O$_{\mathrm{RL}}$&\multicolumn{2}{c}{8.53 $\pm$ 0.04}		    &	\multicolumn{2}{c}{8.60 $\pm$ 0.05}	& \multicolumn{2}{c}{8.57 $\pm $ 0.05}	& \multicolumn{2}{c}{8.65 $\pm$ 0.06} \\
S	& 6.98 $\pm$ 0.04 & 7.09 $\pm$ 	0.05    & 6.98 $\pm$ 0.06 & 7.16 $\pm$ 0.07 	& 7.00 $\pm$ 0.06 & 7.19 $\pm$ 0.07 	& 7.00 $\pm$ 0.07 & 7.30 $\pm$ 0.08 \\
He$_{\mathrm{RL}}$& 10.950 $\pm$ 0.005	& 10.94 $\pm$ 0.01 & 10.972 $\pm$ 0.007	& 10.96 $\pm$ 0.01	&	10.973 $\pm$ 0.007 	&	10.96 $\pm$ 0.01 &	10.990 $\pm$ 0.009	& 10.95 $\pm$ 0.01					\\
\bottomrule
\end{tabular}
\end{table}

\end{landscape}

\end{document}